This preprint is under review. Upon acceptance, please cite the final published version.

# Acute Stroke Classification Using Asymmetric Data Standardization and Orthogonal-Focal Loss Optimized CNN-ViT Cross-Attention Fusion: A Multicenter External Validation Study

Tansel Uyar[1*]

[1] Department of Biomedical Engineering, Faculty of Engineering, Başkent University, Ankara, 06790, Türkiye
***Corresponding author:*** *Assist. Prof. Tansel Uyar, email: tuyar@baskent.edu.tr*

## Abstract

**Background/Objectives:** Rapid differentiation of ischemic and hemorrhagic stroke subtypes is vital for acute management. Although current deep learning models achieve high single-dataset accuracy, they suffer from "shortcut learning" and domain shift in external validation. This study proposes a hierarchical computer-aided diag-nostic (CAD) system mimicking emergency triage. **Methods:** We introduce a 3-channel Radiology-Focused Preprocessing pipeline that feeds radiological Hounsfield Unit (HU) windows into Convolutional Neural Network (CNN) and Vision Transformer (ViT) backbones via asymmetric formats (.mat and .png). The architecture utilizes a novel joint cross-attention fusion network optimized by an Integrated Orthogo-nal-Focal Loss. **Results:** Our system achieved a 95.48% F1-score on the internal stroke dataset. However, multi-center external validation (AISD, BHSD, PhysioNet) using a Mask Anchor filter revealed a drop to the 58-62% range, exposing the "internal validation illusion" caused by hardware noise. Critically, models exhibited high precision (77%) but low sensitivity (57%) for ischemic stroke, proving prone to hemorrhage overdiagnosis. Multi-seed variance analysis demonstrated that the DenseNet201+ViT backbone yielded the lowest deviation (±0.59) and highest stability, making it the most reliable architecture for real-world clinical integration. **Conclusions:** To overcome diagnostic challenges, future research must focus on multimodal architectures that combine 3D volumetric extraction with clinical data.

## 1. Introduction

Cerebral infarction is a major cause of mortality and persistent physical impairment on a global scale. In the context of acute stroke management, the expeditious differentiation of ischemic (obstruction-related) and hemorrhagic (bleeding-related) subtypes in the emergency department is of paramount importance in clinical practice. In subsequent treatment processes, especially in narrow treatment windows such as tissue plasminogen activator (tPA) applications, the time-brain relationship plays a critical role in neuronal survival [1,2]. In the context of emergency triage, computed tomography (CT) has emerged as the prevailing imaging modality due to its expeditious nature, cost-effectiveness, and extensive accessibility. However, in cases of hyperacute ischemic strokes, the presence of subtle density variations on CT scans, which can be challenging to discern with the naked eye (a phenomenon known as the "similarity trap"), can result in inter-observer variability [3]. In light of the substantial workload and physician fatigue prevalent in emergency departments, the incorporation of artificial intelligence and deep learning models as dependable "second readers" in diagnostic processes has become imperative in contemporary medical practice [4–7].

Currently, deep learning algorithms are achieving remarkable success in the field of medical image analysis [3,8–10]. Indeed, the extant literature contains powerful artificial intelligence architectures that have successfully demonstrated their capacity for generalization without overfitting using independent external validation datasets. Nevertheless, contemporary computer-aided diagnostic (CAD) systems in the domain of stroke diagnosis exhibit substantial methodological limitations that must be addressed, including the integration of domain knowledge into the algorithm, the development of classification strategies that emulate clinical workflows, and the implementation of architectural fusion mechanisms. The present study addresses these requirements by focusing on three primary problem areas.

Firstly, there is the destruction of radiological physics in the modeling of medical images. A significant body of research in this field has directly converted cranial CT slices in DICOM format to 8-bit standard PNG or grayscale formats. This process effectively eliminates the Hounsfield Unit (HU) information, which represents the X-ray attenuation coefficients of tissues [6,11,12]. However, radiologically, bone structures exhibit high densities (+1000 HU), hemorrhage areas demonstrate hyperdense densities (+30 to +90 HU), normal brain parenchyma registers at approximately +30 HU, and ischemic lesions manifest as hypodense structures (0 to +30 HU). When raw or standard RGB data is directly fed into classification networks, the models fall into the "shortcut learning" error by focusing on lesion mimics such as skull bone [2,4,13]. The elimination of densities outside the physical boundaries of hemorrhage and ischemia pathologies, in conjunction with the narrowing of the search space through the utilization of HU windows, engenders a substantial multiplier effect in clinical accuracy rates.

Secondly, there are deficiencies in the extent to which deep learning models emulate clinical decision-making processes. A significant number of developed algorithms aim to directly classify cranial CT images into three categories: normal, ischemia, and hemorrhage. However, the phenomenon of "class imbalance," characterized by the disproportionate predominance of normal cases, is frequently observed in stroke datasets. This imbalance has the potential to result in hazardous false negative rates by guiding models toward the healthy class [14]. In clinical practice, however, the physician's initial step is to detect the presence of a stroke lesion (anomaly) in the brain, and the subsequent step is to determine the subtype of the existing stroke. The extant literature has demonstrated that cascaded/hierarchical approaches, which first differentiate between the presence and absence of stroke and then classify stroke types, directly improve diagnostic performance [5,13]. The present study aims to address this research gap by investigating the behavior of cascaded classification strategies in hybrid deep learning architectures, with a particular focus on the limitations of external validation in this context.

The third issue addresses the limitations of unique architectures and the inadequacy of weak evaluation methodologies. Convolutional Neural Networks (CNNs) have demonstrated efficacy in detecting focal hemorrhages with local sensitivity. However, Vision Transformers (ViTs) have shown superiority in modeling the global anatomical context. While hybrid networks that integrate these two architectures have been documented, they typically utilize rudimentary fusion methodologies, such as straightforward prediction fusion or elementary sequential integration of features [9,10]. Moreover, extant literature reveals that the 99% success rates achieved by models trained with a single dataset in a laboratory setting decline significantly (e.g., to levels as low as 78%) when using external clinical datasets with hardware or population differences [15]. There is a need for attentional mechanisms that perform dynamic synthesis in the feature space and transparent, multi-seed-based validation protocols that will demonstrate algorithmic robustness.

In this context, the present study proposes a cascaded deep learning pipeline integrating radiological data, clinical workflows, and hybrid architectures in the diagnosis of acute stroke. The original contributions of the research to the existing literature are as follows:

- **Radiology-Focused Asymmetric and Multi-Channel Standardization:** A distinctive 3-channel preprocessing architecture has been developed, integrating ischemic, hemorrhagic, and normal brain parenchyma windows based on Hounsfield Unit (HU) physics into a unified tensor. In addition, the 3-channel structure has been exported in two distinct data formats, each tailored to the architectural specifications of the respective deep learning network. Specifically, lossless raw tensor matrices (.mat) have been utilized to maintain the numerical precision of the CNN backbones, while image files (.png) have been employed for ViT's spatial patching operation. This has led to the creation of an asymmetric data pipeline, a novel concept that has not been previously documented in the extant literature.
- **Hierarchical Classification Strategy:** To emulate the emergency department triage process, the algorithm is initially trained to detect anomalies (Stroke / No stroke), and subsequently designed in two clinical phases (cascade) to differentiate between ischemic and hemorrhagic subtypes in cases where pathology is detected.

- **Joint Cross-Attention Fusion Architecture:** A novel hybrid network architecture has been developed that synthesizes the local feature extraction power of four distinct CNN backbones (DenseNet201, ResNet50, InceptionResNetV2, EfficientNet-b0) with the global context modeling capability of the ViT architecture by applying a dynamic attention mask in the feature space.
- **Integrated Orthogonal-Focal Loss Function:** In contrast to the cross-entropy approaches that are conventionally utilized in the extant literature, a novel loss function has been devised to enhance the training dynamics of the hybrid network. This function is a mathematical synthesis of two other metrics: the Focused Loss metric, which manages class imbalance and challenging ischemic cases, and the Orthogonal Loss (Cosine Similarity Penalty) metric, the purpose of which is to prevent the feature vectors of the CNN and ViT from memorizing the same information, forcing them to learn complementary features.
- **Analysis of Variance and Detection of External Performance Limitations:** The performance of the selected champion models was transparently evaluated using 5-fold cross-validation and multi-seed analysis of variance. Despite the high success rates in stroke / no stroke detection in the training set, bottlenecks encountered in differentiating ischemic / hemorrhagic in an independent external dataset (external validation), a clinical challenge, were objectively reported. These results revealed the limitations of the model's real-world generalizability.

## 2. Related Works

The utilization of deep learning methodologies for the purpose of stroke diagnosis from cranial non-contrast CT (NCCT) images has witnessed a considerable surge in recent years. A review of the extant literature reveals three predominant themes in classification: data preprocessing strategies, architectural designs, and clinical generalizability (external validation) of models [4–7].

### 2.1. Pre-processing of data and the provision of radiological field information

A significant number of studies in the relevant literature process CT slices in a direct manner in standard 8-bit PNG format, treating pixels as standard visual data. For instance, high classification success rates were reported in studies conducted by Çınar et al. and with the StrokeNeXt model developed by Ekingen et al., who used PNG format data directly [6,11,12]. In a similar vein, Adablanu et al. employed Focal Loss on PNG images to address the issue of class imbalance [14]. However, the failure to remove structures with a value of +1000 HU (Hounsfield Units), such as skull bone, can lead to the models focusing on bone geometry instead of lesion, resulting in a "shortcut learning" error. In order to overcome this limitation, Koska et al. employed radiological field information in order to narrow the search space by eliminating impossible pixels and the skull, thereby increasing the accuracy of a basic model from 74% to 91% [4]. In a similar vein, Uygun and Demirhan, and Karataş et al. have demonstrated that skull-stripping significantly enhances diagnostic accuracy and segmentation Dice scores in studies conducted by DeepBrain [5,13].

### 2.2. Single Architectures and Hybrid Approaches

A plethora of configurations of Convolutional Neural Networks (CNNs) are utilized in classification tasks. Yanar et al. demonstrated the efficacy of the lightweight MobileNetV4 architecture for emergency departments that are constrained in terms of resources [16]. Conversely, Yelken and Ceylan reported that fine ischemic lesion boundaries were more efficiently learned as depth increased in ResNet backbones [17]. The advent of Vision Transformers (ViT) in the field has prompted significant advancements, as evidenced by Katar et al.'s successful experimentation with the ViT architecture in the binary classification of stroke / no stroke [18]. In recent years, hybrid networks that seek to combine the local sensitivity of CNNs with the global context of ViTs have become increasingly prominent. Aziz et al. (EffNet-ViT) attained high accuracy in the identification of ischemic lesions by integrating EfficientNet with ViT [10]. In contrast, Hossen et al. attained consistent success in external validation sets by integrating channel attention and grouped query attention with their hybrid model, termed StroQ-Net [9].

### 2.3. External Validation and Clinical Generalizability

The most significant challenge in the clinical integration of deep learning models is that the 99% accuracy achieved in the laboratory environment is substantially reduced in independent external validation sets from different hardware and patient populations. A survey of the extant literature reveals that only a small number of studies have successfully navigated this domain shift. The DREAM framework proposed by Li et al. achieved 89.6% zero-shot accuracy in an external clinical set with a semi-supervised learning strategy, thus demonstrating an expansion of the limits of generalizability [8]. In a separate study, known as NadHed.AI, a multimodal structure was established by incorporating clinical symptom weights into the features derived from CT images [3]. This approach resulted in an enhanced accuracy of 94.2% in the external cohort test. Diaz et al. (AGTS-Net) reported 90% accuracy in a completely different dataset (Kaggle) with a two-stage anatomically guided cascade architecture [19].

A review of the extant literature indicates the presence of a high degree of potential in the field of cranial CT analysis. However, the clinical community continues to express a need for an integrated methodology that preserves radiological HU information in asymmetric formats suitable for the mathematics of each architecture. Such a methodology would need to dynamically synthesize features with a joint cross-attention mechanism and transparently reveal the challenges of hierarchical classification through external tests. To provide a comprehensive overview of the current state of research in this field, a systematic review of the extant literature was conducted. This review included a meta-analysis of the thematic approaches, data formats, preferred deep learning architectures, and external validation metrics discussed above. The results of this meta-analysis are summarized in Table 1, which also includes a detailed description of the current prominent studies in the field of stroke diagnosis from cranial CT images.

**Table 1.** A thorough examination of contemporary deep learning studies that have achieved prominence in the field of acute stroke diagnosis is conducted, encompassing a comprehensive evaluation of data format, architectural structure, and external generalizability

| Study / Author(s) | Models / Architectures | Data Formatting and Preprocessing | External Validation | Key Findings and Original Contributions |
|---|---|---|---|---|
| Diaz et al., 2026, [19] | AGTS-Net (M* Router) | DICOM / Cranial anatomical cascade | Available (Kaggle NCCT) | The M* external guide crossed anatomical variations with 90% accuracy in the external test. |
| Li et al., 2026, [8] | DREAM (ResNet50 + SSL) | Semi-audited with only 30% label | Available (There are 11,905 sections) | It achieved 89.6% accuracy in the zero-shot protocol. |
| NadHed.AI, 2025, [3] | EfficientNetV2-L + Late Fusion | LAB CLAHE Filter and Clinical Symptoms | Available (There are 200 emergency cases) | Symptom integration increased the accuracy rate to 94.2%. |
| Hossen et al., 2025, [9] | StroQ -Net | Convolutional STEM, Channel Attention, and GQA | Available (Independent dataset) | With its lightweight architecture and 1.10M parameters, it achieved 93.82% accuracy in external testing. |
| Aziz et al., 2026, [10] | EffNet-ViT | 40-120 HU windowing for ischemia | Available (1,045 held-out sections) | It demonstrated 97% stability in internal and external testing without patching. |
| Koska et al., 2024 [4] | MobileNetV2 / EfficientNet-B0 | Raw DICOM and Otsu Thresholding with Skull-stripping | None | Eliminating impossible pixels increased accuracy from 74% to 91%. |
| Uygun and Demirhan, 2025, [13] | Goal-Oriented (Cascade) Segmentation | Skull-stripping and Clinical Normalization | None | First, diagnosis followed by segmentation increased the ischemic Dice score by 19%. |

| Study / Author(s) | Models / Architectures | Data Formatting and Preprocessing | External Validation | Key Findings and Original Contributions |
|---|---|---|---|---|
| Karataş et al., 2022, [5] | DeepBrain (ResNet50 / Inception-v3) | Morphological masking and ROI extraction | None | It is integrated into the desktop application with 99.72% accuracy in detecting stroke presence/absence. |
| Ekingen et al., 2025, [11] | StrokeNeXt | Ready-made 512x512 PNG format | None | 98.8% of F1 score in stroke type differentiation. |
| Adablanu et al., 2025, [14] | Xception | Ready-made PNG format and Focal Loss | None | Focal Loss has eliminated the false negative rate by resolving class imbalance. |
| Uzun and Okuyar, 2025, [20] | YOLOv9-Seg | Original PNG and Mask PNG | None | With a web-based GUI, it achieves 99.71%. mAP has demonstrated segmentation success. |
| Çınar et al., 2023, [12] | EfficientNet-B0 + SVM | PNG grayscale and sharpening | None | 95.13% accuracy by preventing overfitting in shallow data. |
| Satsangi and Narayan, 2026, [21] | Swin, CvT, ResNet Community | QCA and QSA (Query-Attention) Based Fusion | None | The JSC Confidence Score ensures the explainability of decisions. |
| Aksoy et al., 2025,[22] | ResNet-18 | Two-stage fine-tuning and data replication | None | XRAI analysis demonstrated that the model focused on the skull rather than the pathology. |
| Yanar et al., 2025, [16] | MobileNetV4 | 11 different comparisons of modern architecture | None | Heavy Vit's While overfitting occurred, the lightweight CNN (98.42%) stood out in the limited data. |

| Study / Author(s) | Models / Architectures | Data Formatting and Preprocessing | External Validation | Key Findings and Original Contributions |
| --- | --- | --- | --- | --- |
| Qatar et al., 2023, [18] | ViT -Base | Synthetic data enhancement | None | Accuracy in binary classification has been increased to 98.75%. |

## 3. Materials and Methods

### 3.1. Data Sets and Cohort Design

To optimize the proposed two-phase (hierarchical) deep learning pipeline and demonstrate the generalizability of decision boundaries, a comprehensive cohort design consisting of both internal and multi-center external data pools was constructed in this study. Due to the nature of the hierarchical classification architecture, the model's testing and validation strategy was also carried out with two different test datasets, each specific to the clinical objective of each phase.

#### 3.1.1. Internal Training and Cross-Validation Cohort

The primary training set for the TEKNOFEST 2021 Stroke Dataset, which was compiled from the Turkish Ministry of Health's e-Nabız and Teleradiology infrastructure, was utilized to ascertain the weighting of the model's initial phase (Stroke / No Stroke) and subsequent phase (Ischemic / Hemorrhagic) (Figure 1) [23]. The primary cohort under consideration contains a total of 6,650 2D axial CT slices, of which 4,427 are classified as normal (non-stroke), 1,130 are ischemic stroke, and 1,093 are hemorrhagic stroke images. This dataset encompasses a range of real-world clinical data acquired from diverse emergency departments and scanning equipment, exhibiting varying resolutions. The internal cohort was instrumental in establishing the model's structural learning dynamics through its incorporation into stratified five-fold cross-validation and hyperparameter optimization (Bayesian Optimization) processes.

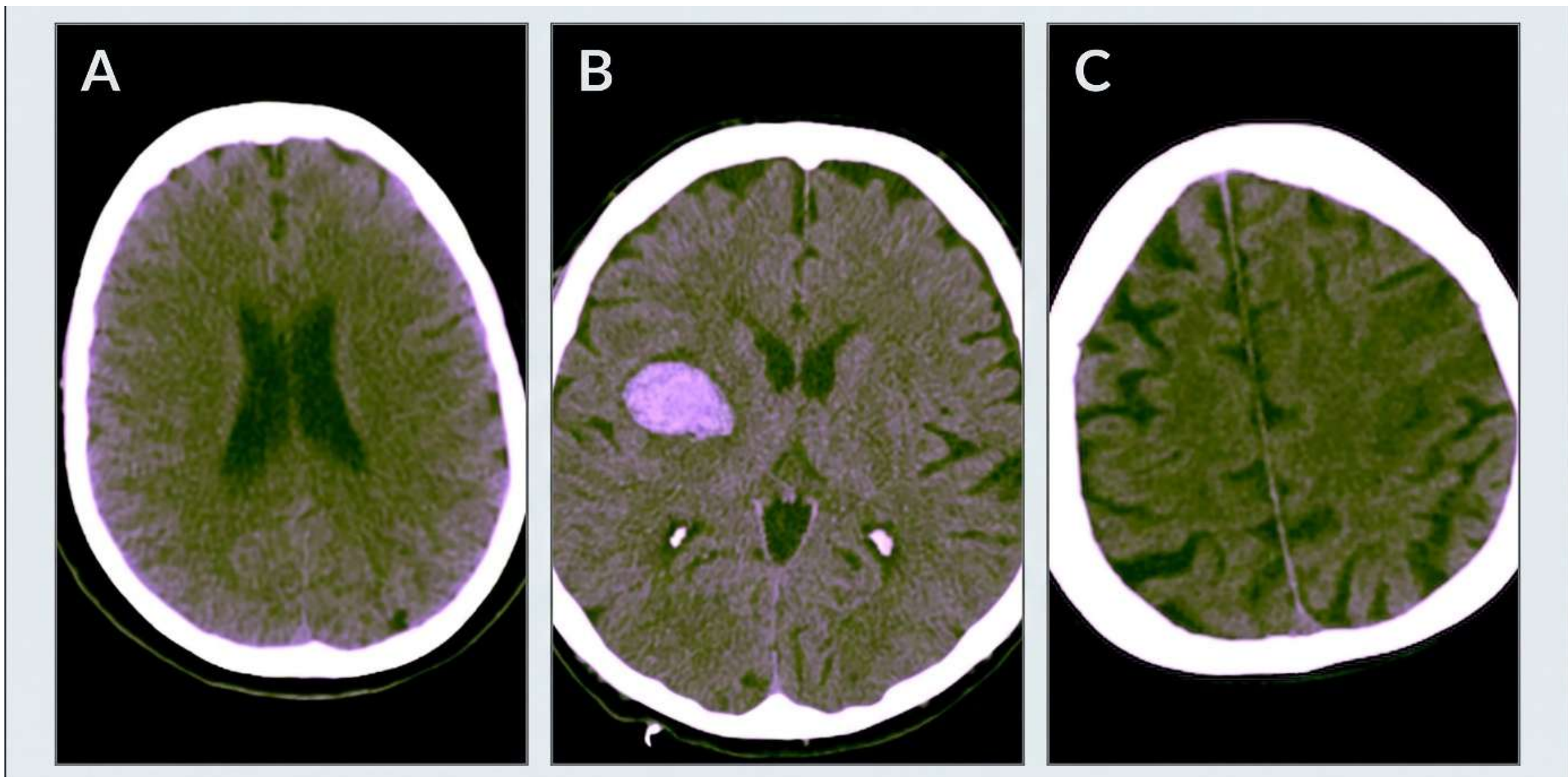

**Figure 1.** Examples of classes from the training dataset, (A) Normal (No Stroke) (B) Hemorrhagic Stroke (Hemorrhage) (C) Ischemic Stroke (Ischemia)

### 3.1.2. Hierarchical Testing and Validation Cohorts

In order to assess whether the model exhibited overfitting to the distributional characteristics of the data in the training set, the model's performance was evaluated using two separate isolated datasets.

- First Stage Test (Stroke / No Stroke Classification): To assess the algorithm's capacity for anomaly detection, images from the TEKNOFEST competition's "first stage," which were not incorporated into the model's training cycles (i.e., isolated), were utilized [23]. The pool, which was designed as a held-out test set, consists of a total of 200 images. The sample included 130 individuals with no history of stroke, as well as 70 individuals with documented stroke pathology. This set is employed to assess the model's proficiency in distinguishing between healthy tissue and anomalies.
- Second Stage Test (Ischemic / Hemorrhagic Classification): To establish the clinical boundaries of subtype differentiation in a transparent manner, the most challenging task of the phased architecture, a "Multi-Center External Validation" cohort was constructed, ensuring complete independence from the training population. A total of 1,874 independent slices were utilized in the study. A total of 1,190 images were retrieved from the AISD (Acute Ischemic Stroke Dataset) collection, which pertains to acute ischemic stroke cases [24]. An additional 684 images were obtained from the BHSD (Brain Hemorrhage Segmentation Dataset) [25] and PhysioNet CT-ICH datasets [26], focusing on intracranial hemorrhage lesions. This configuration presents an environment of domain shift in terms of scanner protocols, device noise variations, and patient demographics, thereby testing the true clinical robustness of the model.

The numerical data from all datasets utilized in the study are outlined in Table 2.

**Table 2.** Class (pathology) based case and cross-sectional distributions of internal and external datasets used in the study

| Data Set (Cohort) | Role / Segment | No Stroke (Normal) | Ischemic Stroke (Ischemia) | Hemorrhagic Stroke (Hemorrhage) |
|---|---|---|---|---|
| TEKNOFEST (Internal) | Train (Stroke / No Stroke) | 3541 | 1778 | |
| | Validation (Stroke / No Stroke) | 886 | 445 | |
| | Train (Ischemic / Hemorrhagic) | N/A | 904 | 874 |
| | Validation (Ischemic / Hemorrhagic) | N/A | 226 | 219 |

<table>
<tr><th>Data Set (Cohort)</th><th>Role / Segment</th><th>No Stroke (Normal)</th><th>Ischemic Stroke (Ischemia)</th><th>Hemorrhagic Stroke (Hemorrhage)</th></tr>
<tr><td></td><td>Internal Test (First Competition Stage)</td><td>130</td><td colspan="2">70</td></tr>
<tr><td>BHSD & PhysioNet (External)</td><td>External Validation</td><td>N/A</td><td>N/A</td><td>684</td></tr>
<tr><td>AISD (External)</td><td>External Validation</td><td>N/A</td><td>1190</td><td>N/A</td></tr>
</table>

#### 3.1.3. Mask Anchor Isolation for External Validation Sets

Given that our gradual classification model generates diagnoses from 2D axial sections, it was necessary to standardize the 3D volumetric scans in the external datasets (AISD, BHSD, PhysioNet) used in the second stage of testing. It is important to note that not every section of a cranial CT scan of a stroke patient contains a pathological lesion. The incorporation of sections devoid of pathology in the external validation test set, designated as "ischemia" or "hemorrhage," results in "Label Noise," thereby inducing a substantial deviation in the evaluation metrics.

To overcome this challenge and ensure methodological transparency (equitable testing conditions), a distinctive "Mask Anchor" algorithm has been incorporated into the prevailing testing procedures documented in the extant literature (Figure 2). The algorithm utilizes segmentation masks delineated by expert radiologists in external datasets as a spatial reference (anchor). The system autonomously filters only 2D axial sections within a 3D volume where the pixel density of the mask (ischemia or hemorrhage) is greater than zero. This indicates that the lesion is radiologically visible and definitive. This approach enables the subsequent external testing process, comprising 1,190 ischemic and 684 hemorrhagic sections, to refine the model's classification capability to measure exclusively clinical sections containing definitive pathology.

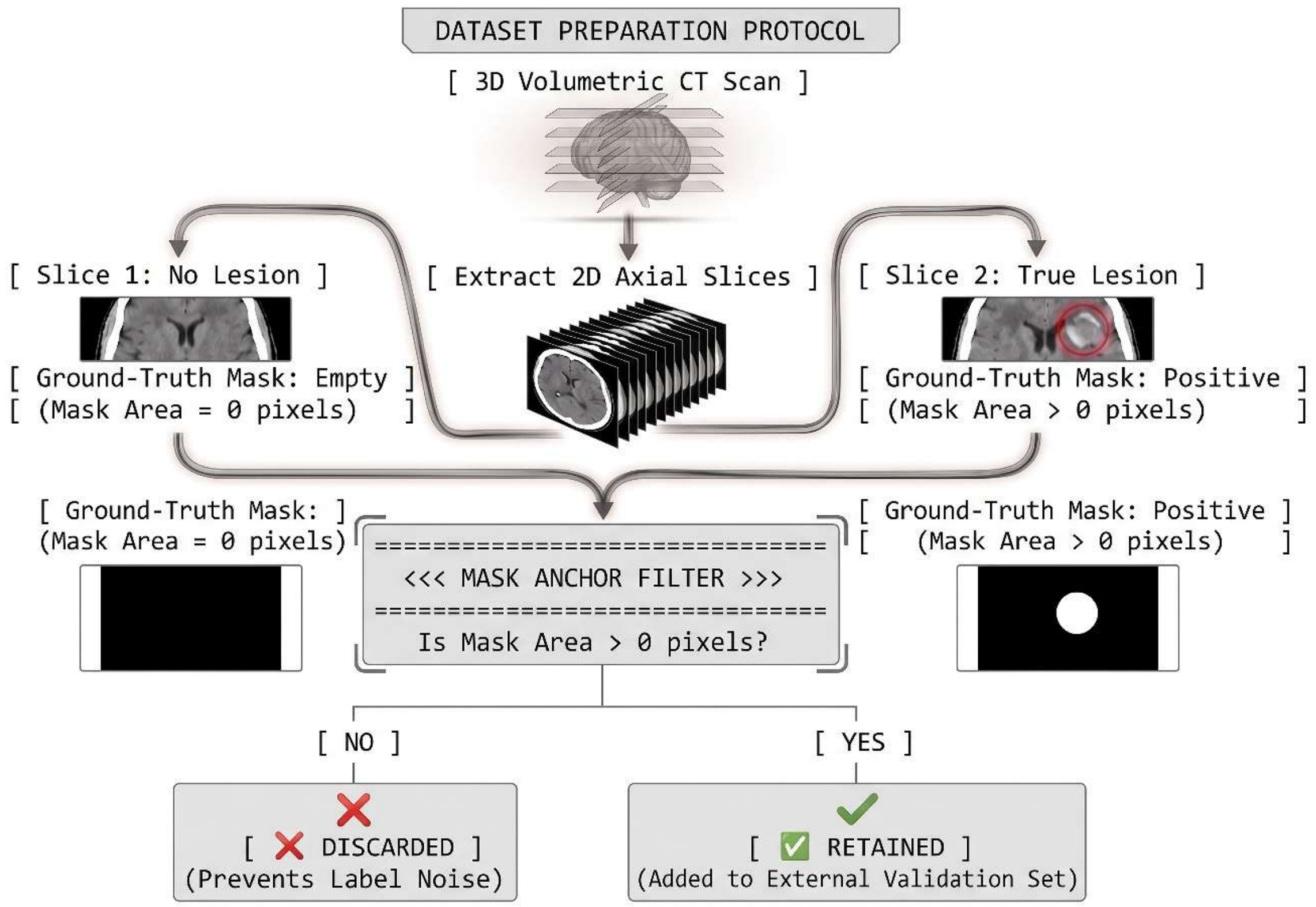

**Figure 2.** Visualization of the Mask Anchor algorithm applied in the external validation cohort; elimination of pathology-free sections and isolation of lesioned sections confirmed only with a radiologist's mask

### 3.2. Radiology-Focused Multichannel Preprocessing and Asymmetric Data Optimization

The efficacy of AI-based diagnostic models in acute stroke diagnosis is contingent upon the quality of the data provided to the system and the incorporation of cranial area-specific radiological boundaries into the algorithm. A significant body of research in this field has involved the direct conversion of DICOM format CT slices to standard images, a process that often results in the loss of valuable pixel density data (HU values). In this study, a radiology-focused multi-channel preprocessing and asymmetric data feed pipeline was developed to highlight the subtle structure of stroke lesions and prevent models from falling into shortcut learning errors.

#### 3.2.1. Radiological Multi-Windowing and Tensor Construction

In the initial phase of the processing pipeline, signals derived from raw DICOM data were transformed into physical Hounsfield Unit (HU) values using linear rescaling (Rescale Slope and Intercept). Subsequently, three distinct radiological windows (Level and Width) were delineated to differentiate between varying tissue densities:

- Channel 1 (General Anatomy / Brain Window): The center level is set at 40 HU, and the width is set at 80 HU, thus highlighting the general brain parenchyma and gray/white matter structure.
- Channel 2 (hemorrhagic/subdural window): The center level is set at 50 HU, and the width is set at 150 HU. This is done in an effort to cover hyperdense lesion areas associated with acute hemorrhage.
- Channel 3 (Ischemic/Narrow Stroke Window): The center level is set at 40 HU, and the width is set at 40 HU. This configuration compresses the area into a very narrow range, thereby highlighting subtle hypodensity differences, particularly those seen in ischemic strokes.

The algorithm calculates the lower (minHU) and upper (maxHU) limits for each of these windows. It then mathematically trims impossible pixels outside the limits (e.g., a skull bone with a +1000 HU value) and normalizes all densities between 0 and 1. These three normalized channels are then superimposed to form a single multi-channel synthesized tensor.

### 3.2.2. Asymmetric Data Output for Architectural Requirements

A notable methodological contribution of this study to the existing literature is the introduction of a multi-channel tensor that is not presented to the CNN and ViT arms of the fusion network in the same format. Due to the disparate mathematical principles underlying feature extraction in these networks, the data feeding process is structured to be "asymmetric" in both structure and dimension.

- Numerical Accuracy for CNN Backbones (.mat Format): The tensor generated for Convolutional Neural Networks (DenseNet, ResNet, etc.) has been rescaled to 224 x 224 pixel dimensions. To enable CNN to prioritize radiological density mathematics, the data was exported in its original floating-point accuracy in ".mat (MATLAB Matrix)" format without any lossy compression (quantization loss).
- Spatial Optimization for ViT Architecture (.png Format): ViTs necessitate a broader global context due to their processing of data in spatial patches. Consequently, the tensor has been scaled up to 384 x 384 to ensure compatibility with the ViT input layer. Subsequently, the mathematical values were multiplied by 255 to convert them to 8-bit unsigned integer (uint8) format and saved as a standard 3-channel RGB .png image.

## 3.3. Joint Cross-Attention Fusion Architecture

The hierarchical diagnostic mechanism targeted in this study is designed to simulate emergency department triage by first classifying stroke as "Stroke / No Stroke" and then differentiating between "Ischemic/Hemorrhagic" subtypes in positive cases. In order to successfully execute this challenging clinical task, a "Joint Cross-Attention Fusion Architecture" has been developed. This architecture employs the hierarchical local sensitivity of CNNs and the global spatial context of ViTs simultaneously.

The system consists of two modular steps (Figure 3): Train the base networks (Stage 1) and cross-attention fusion (Stage 2).

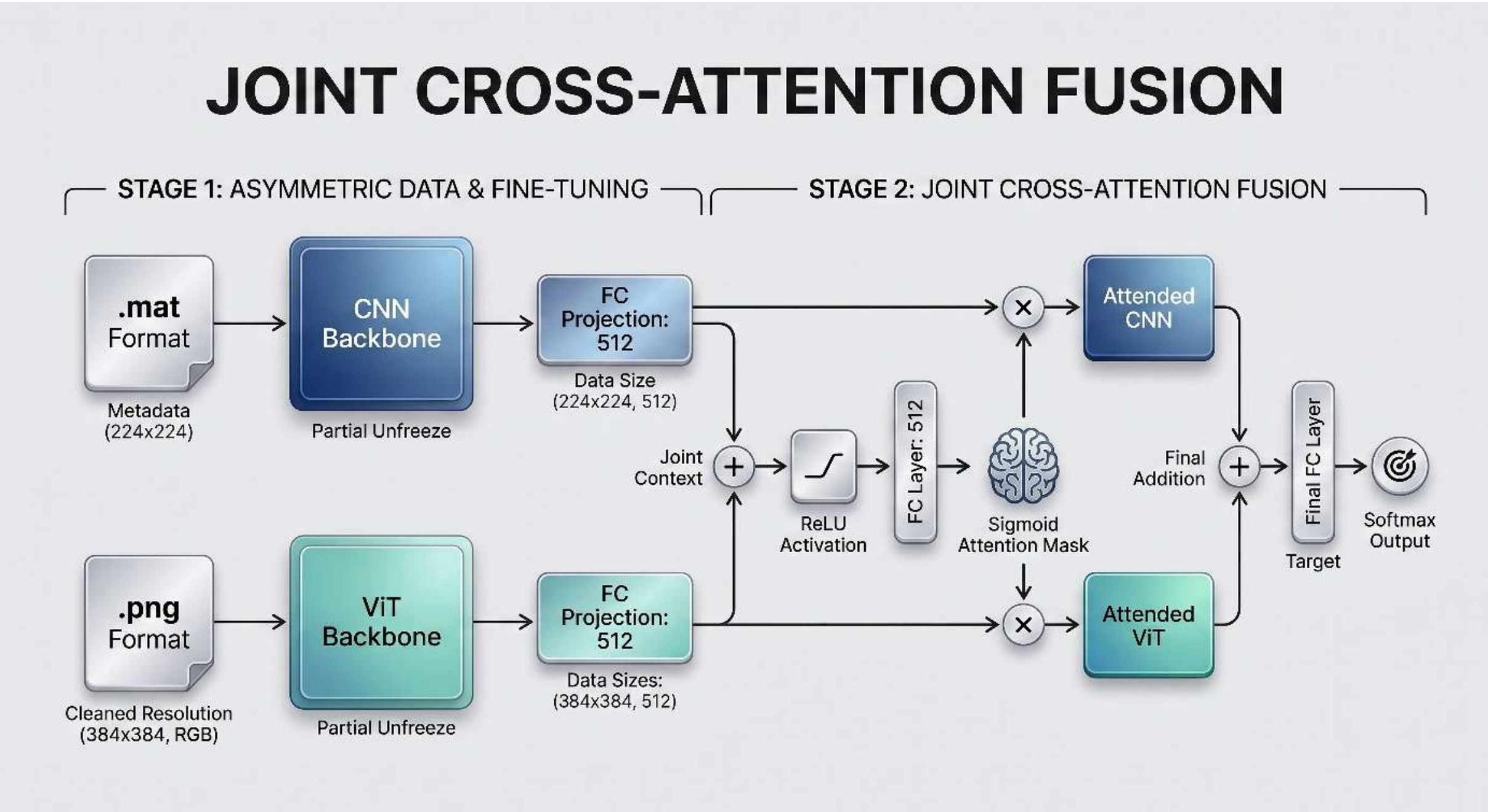

**Figure 3.** Structural architecture of the proposed deep learning pipeline

### 3.3.1. Construction of Asymmetric Backbone Networks and Partial Unfreeze

In order to provide robust features for the fusion mechanism, four different CNN backbones (DenseNet201, ResNet50, EfficientNet-b0, InceptionResNetV2) and a ViT architecture (base-16-imagenet-384), both proven successful in the literature, were tested. To avoid the "catastrophic forgetting" problem by distorting the overall anatomical weights of the medical images presented to the network, a "partial unfreezing" strategy was applied. In the CNN backbones, the last 15% of learnable layers were opened for training, while in the ViT architecture, only the last 3 blocks of the 12 main blocks were opened for training. These networks were fine-tuned asymmetrically with first classification step (Stroke / No Stroke) and second classification step (Ischemic/Hemorrhagic) labels over 10 epochs at a low learning rate ($1 \times 10^{-4}$) using SGDM optimization.

### 3.3.2. Cross-Attention and Joint Context Synthesis

Feature vectors of size 1 x 512 were extracted from the finely tuned CNN and ViT networks. However, direct concatenation of these raw vectors creates semantic noise. To dynamically synthesize the features, a "Gated Attention Mask," rarely found in the literature, was constructed:

- Dynamic Projection: Vectors from CNN and ViT are projected onto a 512-dimensional shared feature space via Fully Connected Layers.

- Extracting the Joint Context: 512-dimensional projections of two different architectures were element-wise aggregated and then subjected to ReLU activation to create a "joint context".
- Sigmoid Masking and Fusion: The joint context was passed through a new 512-layer structure and compressed between the Sigmoid function [0, 1] to obtain a dynamic Attention Mask. This mask was independently multiplied by the features of both base networks (CNN and ViT) to produce the “Attended CNN” and “Attended ViT” features.

The final features that survived the cross-masking process were collected in a final step and passed to the Softmax layer to decide on the relevant diagnostic stage (Stroke / No Stroke or Ischemic / Hemorrhagic). Thanks to this design, for instance, in a group of ischemic pixels where the CNN network is mistaken, the spatial attention of the ViT network can intervene and correct the final decision.

### 3.4. Integrated Orthogonal-Focal Loss Function

In order to optimize the training dynamics of the proposed Joint Cross-Attention fusion architecture and to prevent the hybrid network from falling into the "feature homogenization" trap, a three-component integrated loss function has been designed. This function differs from standard loss metrics in literature. This function, which was developed for the specific purpose of addressing these issues, has three distinct components. The first component is responsible for monitoring class imbalance. The second component is responsible for monitoring inter-architecture complementarity. The third component is responsible for monitoring model generalizability.

#### 3.4.1. Focal Loss for Class Imbalance

The fact that normal cases outnumber ischemic and hemorrhagic stroke cases in clinical datasets causes the standard Cross-Entropy function to bias towards the majority class. Furthermore, the subtle hypodensity created by ischemic stroke lesions on CT is a difficult problem to diagnose for the model. To overcome these challenges, Focal Loss has been integrated into the system.

$$Loss_{Focal} = -\frac{1}{N}\sum_{i=1}^{N}(1-p_t)^{\gamma}\log(p_t) \tag{1}$$

Here, $p_t$ represents the probability value predicted by the model for the correct class, while gamma ($\gamma$) (focusing parameter) is set to 2.0. This allows the model to shift its mathematical weight towards ischemic cases that are difficult to diagnose, rather than focusing on easily distinguishable hemorrhagic bleeding or healthy tissues.

#### 3.4.2. Orthogonal Loss for Feature Diversity (Cosine Similarity Penalty)

In the context of CNN and ViT arms being trained in tandem within hybrid networks, there exists a considerable probability of co-adaptation, wherein both networks acquire the same predominant texture characteristics. To mathematically enforce complementary feature learning (one for local pixels, the other for global spatial binding), an "Orthogonal Loss (Cosine Similarity Penalty)" has been designed. The absolute cosine similarity between the 512-dimensional CNN ($f_{CNN}$) and ViT ($f_{ViT}$) feature vectors obtained from the projection layers is calculated and added to the loss function as a penalty.

$$Loss_{Orth} = \frac{1}{N}\sum_{i=1}^{N}\frac{|\mathbf{f}_{CNN}\cdot\mathbf{f}_{ViT}|}{||\mathbf{f}_{CNN}||||\mathbf{f}_{ViT}||} \tag{2}$$

This orthogonal approach compels networks to learn distinct, independent (orthogonal) representations to minimize error rate (loss), thereby maximizing the information entropy of the decision-making mechanism.

#### 3.4.3. Total Integrated Loss (Total Loss)

To ensure the asymptotic stability of the final model, these two fundamental metrics are combined with L2 Regularization, which controls weight decay. The total loss function governing the overall training cycle is formulated as follows:

$$Loss_{Total} = Loss_{Focal} + \lambda * Loss_{Ortho} + Loss_{L2} \tag{3}$$

The hyperparameter $\lambda$ in the equation balances the effect of the orthogonal penalty on the overall loss, and its optimum value is determined through Bayesian optimization processes.

### 3.5. Experimental Setup and Transparent Evaluation Protocol

To demonstrate the clinical generalizability and statistical stability of the proposed deep learning architecture, training and testing processes were integrated with multi-seed stratified k-fold cross-validation and ensemble strategies. All experimental studies were conducted using an Intel i7-10700 processor, 32 GB of 2,933 MHz DDR4 RAM, and an NVIDIA RTX 4070 graphics card with 12 GB of GDDR6X VRAM. MATLAB 2026a software (MathWorks, Massachusetts, USA) was used for all studies.

#### 3.5.1. Bayesian Hyperparameter Optimization

To optimize the complex learning space of the Joint Cross-Attention Fusion network, Bayesian optimization was utilized in contrast to the conventional trial-and-error (grid search) method [27]. The search space was defined to include the following parameters: The initial learning rate was set between $1 \times 10^{-5}$ and $5 \times 10^{-3}$; the orthogonal loss weight was set between 0.01 and 0.6; the L2 regularization coefficient was set between $10^{-6}$ and $10^{-3}$; the ADAM optimization momentum was set between 0.80 and 0.99; and the batch size was set to 16, 32, or 64. Following a total of 50 successive evaluations, the algorithm determined the optimal hyperparameter configuration, which yielded the lowest validation loss and was designated as the optimal configuration.

### 3.5.2. Multi-Seed and Cross-Validation Process

In order to eliminate the chance factor (bias) that might arise from random initialization of network weights, the training process was repeated three times using three different global seeds (Seed: 42, 123, 2026). The internal dataset for each seed was subjected to five-fold cross-validation, yielding a total of 15 independent fusion models. This approach effectively prevented the model from overfitting to a specific subset of the data [28].

### 3.5.3. Ensemble Testing Strategies for External Validation

The clinical generalizability of the 15 independent models obtained during the training phase in external validation sets (AISD, BHSD, PhysioNet) was tested from two different perspectives using the "Soft Voting" architecture, where the classification probabilities were averaged [29]:

- Strategy 1 - Cross-Seed Best-Model Ensemble: The objective of this strategy is to assess the maximum potential performance of the model within a clinical context. From each seed (42, 123, 2026), only one of the most successful models (champions) that achieved the highest Macro F1-Score in the internal (TEKNOFEST) validation set was selected from among its five folds. The Soft Voting predictions of the three selected independent champion models in the external dataset were combined to generate the final diagnostic performance report. Selecting the models without seeing the external dataset completely eliminated the possibility of optimistic bias (data leakage).
- Strategy 2 - Intra-Seed K-Fold Ensemble Variance Analysis: The objective of this method is to assess the structural stability (i.e., robustness) of the model and its resistance to distributional variance in the training data. The cumulative success of a specific seed was determined by combining the outputs of five distinct fold models from an external dataset using a Soft Voting approach. This process was repeated separately for three different seeds, and the external validation successes were presented in the format "Mean +/- Standard Deviation" to transparently demonstrate the model's confidence interval.

### 3.5.4. Performance Metrics

To objectively evaluate the diagnostic capacity of the model in external medical datasets containing class imbalance (e.g., 1,190 Ischemia, 684 Hemorrhage), standard classification metrics based on the Confusion Matrix were used [30,31]. The complexity matrix components underlying the classification metrics used in this study are defined as follows in the context of acute stroke diagnosis:

True Positive (TP): Correct identification of the relevant stroke pathology by the model.

True Negative (TN): Correct exclusion by the model of sections that do not contain the targeted pathology.

False Positive (FP - Type I Error): Incorrect labeling of healthy tissue as a stroke lesion (false alarm).

False Negative (FN - Type II Error): Completely missing the existing stroke lesion by the model (the most clinically risky error).

$$\text{Accuracy}=\frac{\text{TP+TN}}{\text{TP+TN+FP+FN}} \tag{4}$$

$$\text{Recall}=\frac{\text{TP}}{\text{TP+FN}} \tag{5}$$

$$\text{Precision}=\frac{\text{TP}}{\text{TP+FP}} \tag{6}$$

$$\text{Specificity}=\frac{\text{TN}}{\text{TN+FP}} \tag{7}$$

$$\text{F1-score}=\frac{\text{TP}}{\text{TP+}\frac{1}{2}\text{(FP+FN)}} \tag{8}$$

$$\text{AUC=Area}\left(\frac{\text{Recall}}{\text{1-Specificity}}\right) \tag{9}$$

Following the calculation of the evaluation metrics for each class individually, macro-average scores were obtained by taking the arithmetic mean of all results. This approach was adopted to prevent the manipulation of performance resulting from inter-class data imbalance. Furthermore, the Area Under the Curve (Macro-AUC) metric was incorporated into the evaluation protocol to assess the model's efficacy at various cutoff values [31].

## 4. Results

### 4.1. Phase 1 - Stroke / No Stroke Classification (Ablation Study)

A comprehensive ablation study was conducted on the internal training dataset (TEKNOFEST) to ascertain the most compatible Convolutional Neural Network (CNN) to be integrated into a fixed Vision Transformer (ViT) backbone in the developed Joint Cross-Attention network. In this initial phase, which emulates emergency department triage (gradual classification approach), four distinct CNN backbones (ResNet50, DenseNet201, EfficientNet-b0, and InceptionResNetV2) were evaluated for their capacity to diagnose "Stroke" and "No Stroke" by aligning with the ViT architecture, respectively.

The structural stability of the four distinct hybrid fusion combinations was assessed by measuring their resistance to within-data variance (distributional differences). To this end, three distinct global seeds (Seed: 42, 123, 2026) and stratified 5-fold cross-validation were utilized. The mean and standard deviation of the cumulative success of these 15 independent models tested in the internal validation folds is presented in Table 3.

**Table 3.** Performance comparison of different fusion combinations (CNN + ViT) in the TEKNOFEST dataset for first classification (Stroke / No Stroke) step

| Fusion Architecture | Accuracy (%) | Precision (%) | Recall (%) | Specificity (%) | F1-Score (%) | AUC (%) |
|---|---|---|---|---|---|---|
| ResNet50 + ViT | 94.75 ± 0.68 | 94.29 ± 0.82 | 93.89 ± 0.80 | 93.89 ± 0.80 | 94.08 ± 0.76 | 98.64 ± 0.37 |
| **DenseNet201 + ViT** | **95.88 ± 0.58** | **95.69 ± 0.64** | **95.03 ± 0.80** | **95.03 ± 0.80** | **95.34 ± 0.67** | **98.94 ± 0.28** |
| EfficientNet-b0 + ViT | 95.37 ± 0.56 | 95.09 ± 0.58 | 94.47 ± 0.77 | 94.47 ± 0.77 | 94.77 ± 0.64 | 98.82 ± 0.26 |
| InceptionResNetV2 + ViT | 95.39 ± 0.67 | 95.07 ± 0.79 | 94.55 ± 0.75 | 94.55 ± 0.75 | 94.80 ± 0.75 | 98.80 ± 0.31 |

* Results are presented as Mean ± Standard Deviation over 15 independent models (3 seed x 5-fold).

As demonstrated in Table 3, the DenseNet201 + ViT fusion model exhibited a notable advantage over other combinations in all internal evaluation metrics. The model's low standard deviation of ±0.58 during the training phase is indicative of its resistance to overfitting and its stability across different layers. To more transparently analyze the structural fluctuations (variance) exhibited by four different fusion combinations over 15 independent cycles during the training process, a grouped boxplot was created as shown in Figure 4.

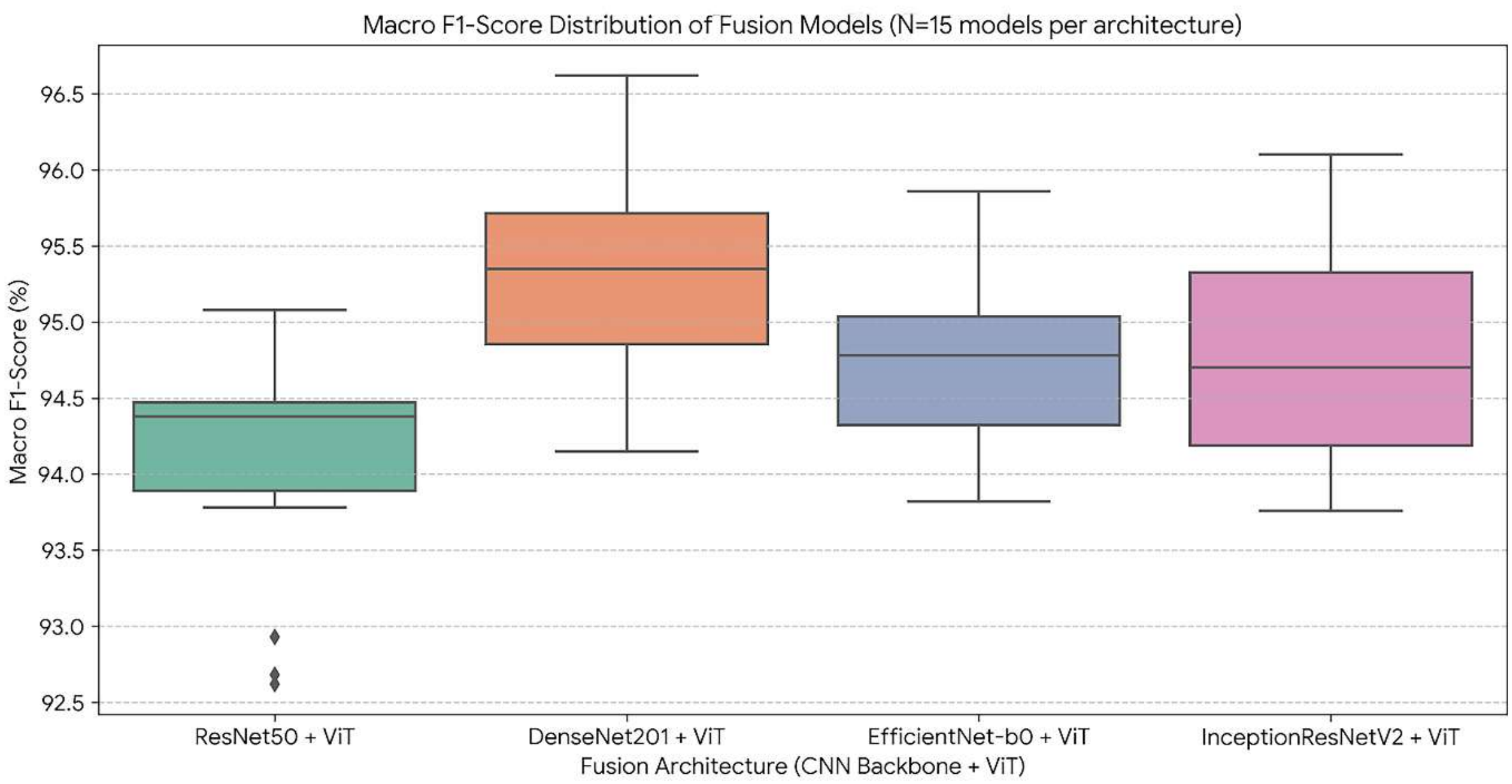

**Figure 4.** Ablation graph shows the structural stability of four different asymmetric fusion combinations

The boxplots display the distribution of Macro F1-Scores across 15 independent training cycles (3 global seeds x 5 folds) evaluated strictly on the internal validation (TEKNOFEST) dataset for first classification (Stroke vs. No Stroke). The DenseNet201 + ViT combination exhibits the highest median score and the tightest interquartile range (IQR), proving its optimal robustness for external deployment. In light of this internal training evidence, the DenseNet201 + ViT structure was selected as the "champion backbone" for application to external validation sets (AISD, BHSD) in the clinical external testing phase of the study (in the Strategy 1 and Strategy 2 ensemble models). However, the results of other model pairings will also be examined within the external validation process.

### 4.2. Phase 2 - Ischemic / Hemorrhagic Classification (Ablation Study)

In the second phase of our hierarchical triage model, after an anomaly (stroke) is detected in a CT scan, the system focuses on determining whether the pathology is ischemic or hemorrhagic, a clinically more challenging task. For this task, the ViT backbone, which successfully captures global features, was used as the basis, and four different architectures (ResNet50, DenseNet201, EfficientNet-b0, InceptionResNetV2) were evaluated to identify the most suitable local feature extractor (CNN) to integrate into this network. To increase the reliability of the analyses and transparently observe the risk of overfitting in different data subsets, 5-fold cross-validation was applied with three independent global seed values on the internal dataset, and 15 independent model cycles were run per architecture. The cumulative performance metrics (Mean ± Standard Deviation) obtained by the fusion algorithms for the second phase are summarized in Table 4.

**Table 4.** Performance comparison of different fusion combinations (CNN + ViT) in the TEKNOFEST dataset for second classification (Ischemic / Hemorrhagic) step

| Fusion Architecture | Accuracy (%) | Precision (%) | Recall (%) | Specificity (%) | F1-Score (%) | AUC (%) |
|---|---|---|---|---|---|---|
| ResNet50 + ViT | 94.34 ± 1.07 | 94.36 ± 1.06 | 94.34 ± 1.08 | 94.34 ± 1.08 | 94.34 ± 1.07 | 98.59 ± 0.24 |
| **DenseNet201 + ViT** | **95.48 ± 0.76** | **95.49 ± 0.77** | **95.48 ± 0.76** | **95.48 ± 0.76** | **95.48 ± 0.76** | **98.70 ± 0.40** |
| EfficientNet-b0 + ViT | 94.61 ± 1.21 | 94.64 ± 1.22 | 94.60 ± 1.21 | 94.60 ± 1.21 | 94.61 ± 1.21 | 98.59 ± 0.47 |
| InceptionResNetV2 + ViT | 94.67 ± 1.58 | 94.69 ± 1.58 | 94.67 ± 1.57 | 94.67 ± 1.57 | 94.67 ± 1.58 | 98.49 ± 0.70 |

* Results are presented as Mean ± Standard Deviation over 15 independent models (3 seed x 5-fold).

An examination of the results summarized in Table 4 reveals that DenseNet201 + ViT fusion architecture maintains its statistical superiority compared to other combinations, even in a radiologically more challenging task such as differentiating ischemia from hemorrhage. The DenseNet201 backbone demonstrated the highest degree of success, achieving 95.49% accuracy and a 95.48% F1-Score. Additionally, it exhibited inter-fold stability, with the lowest standard deviation of ±0.76. This finding suggests that DenseNet's densely connected structure leads to the production of more successful feature maps than other networks in capturing subtle hypodensity differences caused by ischemia. In contrast, the InceptionResNetV2 + ViT architecture demonstrated a competitive average of 94.67% F1-Score.

However, its high standard deviation of ±1.58 indicates that the network exhibits increased vulnerability (higher variance) to initialization weights and fold changes.

The ablation plot, which was created to visualize the structural stability (variance) of the 15 independent models obtained via the Macro F1-Score, is presented in Figure 5.

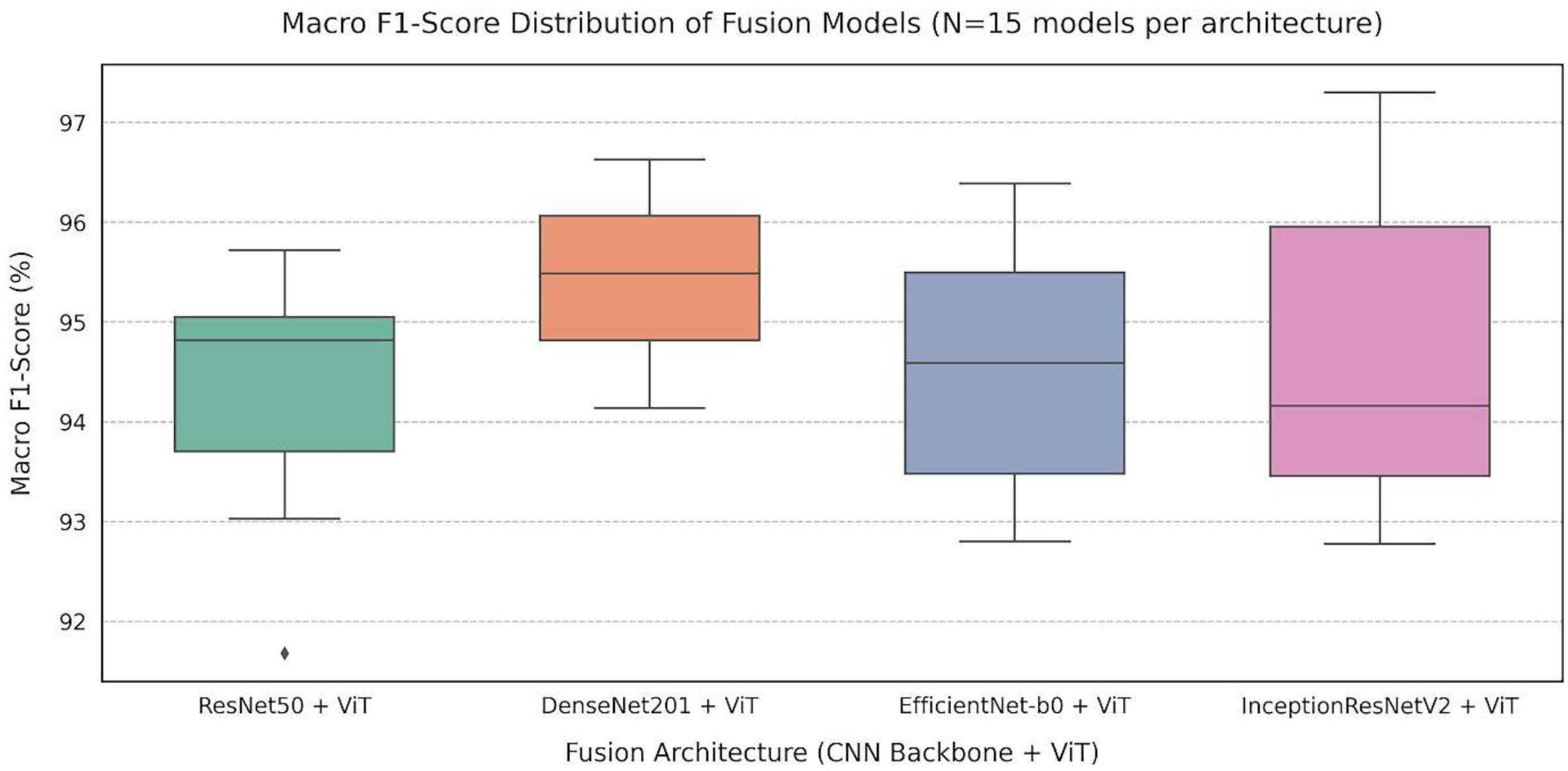

**Figure 5.** Ablation graph showing the structural stability and performance variance (Macro F1-Score distribution) of four different fusion backbones in the second phase (Ischemic / Hemorrhagic) classification task

In boxplot analysis, the DenseNet201 + ViT model was found to have the highest median value (95.49%) and the narrowest interquartile range (IQR: 1.24). In contrast, while the InceptionResNetV2 + ViT architecture achieved the highest singular success across all cycles with the maximum F1-Score (97.30%), it exhibited a significant performance fluctuation between folds and random initiation seeds by producing the widest interquartile range (IQR: 2.49). This statistical distribution validates the structural reliability provided by densely coupled feature maps (DenseNet201) in capturing the subtle radiological boundaries of ischemic tissue, and clearly demonstrates how critical not only mean metric success but also variance stability is in the clinical generalizability of AI models. When the cross-validation findings from the first phase (Stroke / No Stroke) and the second phase (Ischemic / Hemorrhagic) were combined, the DenseNet201 + ViT combination was registered as the 'champion backbone' of the study and selected for independent external validation tests in the next step. However, the results of other model pairings will also be examined within the external validation process.

### 4.3. External Validation and Generalizability Outcomes

The most significant challenge to the integration of artificial intelligence models into clinical practice is the observation that the high success rates achieved in controlled laboratory environments (internal validation sets) experience a substantial decline when applied to external datasets characterized by distinct hardware (e.g., scanners), protocols, and patient demographics (i.e., domain shift). In this study, the objective was to test whether the model falls into the trap of structural overfitting. Strategy-1 (Cross-Seed Best-Model Ensemble) was implemented in the first part of external validation. Strategy-2 (Intra-Seed K-Fold Ensemble Variance Analysis) was implemented in the second part of external validation.

#### 4.3.1. Strategy-1: Cross-Seed Best-Model Ensemble

In this strategy, champion models (with frozen hyperparameters) that received the highest scores from each initial seed in the TEKNOFEST internal dataset were selected and tested in completely isolated external cohorts using a soft voting mechanism. The model's ability to distinguish between healthy brain tissue and stroke pathology was evaluated in an independent test pool of 200 slices that were not included in the training process. The class-based ensemble performance of four different fusion architectures on this dataset is presented in Table 5.

**Table 5.** Performance metrics of fusion architectures under Strategy 1 configuration, broken down by class, in an independent external test set (Stroke / No Stroke)

| Fusion Architecture | Class | Precision (%) | Recall (%) | Specificity (%) | F1-Score (%) | AUC (%) |
|---|---|---|---|---|---|---|
| **ResNet50 + ViT** | No Stroke | 90.00 | 90.00 | 81.43 | 90.00 | 94.43 |
| | Stroke | 81.43 | 81.43 | 90.00 | 81.43 | 94.43 |
| | Macro-Average | 85.71 | 85.71 | 85.71 | 85.71 | 94.43 |
| **DenseNet201 + ViT** | No Stroke | 90.30 | 93.08 | 81.43 | 91.67 | 94.34 |
| | Stroke | 86.36 | 81.43 | 93.08 | 83.82 | 94.34 |
| | Macro-Average | 88.33 | 87.25 | 87.25 | 87.74 | 94.34 |
| **EfficientNet-b0 + ViT** | No Stroke | 91.04 | 93.85 | 82.86 | 92.42 | 94.49 |
| | Stroke | 87.88 | 82.86 | 93.85 | 85.29 | 94.49 |
| | Macro-Average | 89.46 | 88.35 | 88.35 | 88.85 | 94.49 |
| **InceptionResNetV2 + ViT** | No Stroke | 89.63 | 93.08 | 80.00 | 91.32 | 94.73 |
| | Stroke | 86.15 | 80.00 | 93.08 | 82.96 | 94.73 |
| | Macro-Average | 87.89 | 86.53 | 86.53 | 87.14 | 94.73 |

An examination of the class-based distribution in Table 5 reveals that the models' recall sensitivity rates for stroke detection remain in the ~80-82% range on average, while their recognition rates for healthy tissue (no stroke) reach up to ~93%. This asymmetrical structure suggests that the models exhibit a heightened degree of conservatism during the search for anomalies, which can result in the production of false negative results due to the omission of subtle stroke findings. It is noteworthy that the DenseNet201 backbone, which had been identified as the optimal architecture during the internal training phase, has been replaced in this phase by the lightweight EfficientNet-b0 architecture. This replacement has resulted in an average F1-Score of 88.86%. It has been observed that the EfficientNet-b0 architecture, with its reduced number of parameters, exhibits superior capacity for feature generalization across data with diverse distributions.

The most clinically challenging phase of the study, the subtype classification of ischemic and hemorrhagic, was tested on external datasets collected from completely different centers (AISD, BHSD, PhysioNet). The multi-center classification performance of the algorithms is reported in Table 6.

**Table 6.** Strategy 1 (Ensemble) class-based performance findings of the models in the Ischemic and Hemorrhagic subtype classification task on multi-center external datasets

| Fusion Architecture | Class | Precision (%) | Recall (%) | Specificity (%) | F1-Score (%) | AUC (%) |
|---|---|---|---|---|---|---|
| **ResNet50 + ViT** | Ischemic | 77.58 | 57.56 | 71.05 | 66.09 | 70.04 |
| | Hemorrhagic | 49.04 | 71.05 | 57.56 | 58.03 | 70.04 |
| | Macro-Average | 63.30 | 64.30 | 64.30 | 62.05 | 70.04 |
| **DenseNet201 + ViT** | Ischemic | 75.52 | 54.71 | 69.15 | 63.45 | 67.47 |
| | Hemorrhagic | 46.74 | 69.15 | 54.71 | 55.78 | 67.47 |
| | Macro-Average | 61.13 | 61.92 | 61.92 | 59.61 | 67.47 |
| **EfficientNet-b0 + ViT** | Ischemic | 74.31 | 52.02 | 68.71 | 61.20 | 65.71 |
| | Hemorrhagic | 45.15 | 68.71 | 52.02 | 54.49 | 65.71 |
| | Macro-Average | 59.72 | 60.36 | 60.36 | 57.84 | 65.71 |
| **InceptionResNetV2 + ViT** | Ischemic | 74.63 | 54.87 | 67.54 | 63.24 | 66.63 |
| | Hemorrhagic | 46.25 | 67.54 | 54.87 | 54.90 | 66.63 |
| | Macro-Average | 60.43 | 61.20 | 61.20 | 59.07 | 66.63 |

The clinical findings in Table 6 dramatically illustrate the fundamental weakness (domain shift effect) of computer-aided diagnostic (CAD) systems. Algorithms that achieved high scores in the internal validation set showed macro F1 scores dropping to the 58%-62% range in multicenter tests. Examining the class-based data revealed a critical pathological learning vulnerability: All models exhibited very high precision (~74-77%) but very low sensitivity (~52-57%) in the Ischemic class. In contrast, the situation was reversed in the Hemorrhage class; while sensitivity reached 71%, precision remained at 45-49%. This suggests that the models tend to interpret device noise or artifacts in external data as hyperdense (bright) hemorrhage foci, leading to "overdiagnosis." However, when they make an ischemic decision, they mathematically prove that this decision is clinically highly accurate. In this challenging clinical frontier, the ResNet50 + ViT architecture, which can extract deeper hierarchical patterns, has proven to be the most resilient network to domain changes (scanner noise, slice thickness variability) in the external dataset, as it best balances class variances.

#### 4.3.2. Strategy-2: Intra-Seed K-Fold Ensemble Variance Analysis

Strategy-2 was implemented to investigate whether the performance degradation exhibited by AI models on external datasets (domain shift) stemmed from the weight distribution of the selected individual models or from a lack of structural generalizability in the architecture. In this approach, the external test decisions of each global seed's five internal folds were combined using the "Soft Voting" method. The mean and standard deviation values of the resulting ensemble scores on three independent seeds were reported. In this manner, the variance (vulnerability) of the model against the random initialization weights and data distribution was measured transparently.

In the anomaly detection task performed on the external dataset, the statistical distributions of the fold-ensemble results of the fusion models are presented in Table 7.

**Table 7.** Class-based mean and standard deviation performance metrics obtained in the independent external test set (Stroke / No Stroke) under Strategy 2 configuration (N=3 Seeds)

| Fusion Architecture | Class | Precision (%) | Recall (%) | Specificity (%) | F1-Score (%) | AUC (%) |
|---|---|---|---|---|---|---|
| **ResNet50 + ViT** | No Stroke | 90.75 ± 0.45 | 93.08 ± 0.77 | 82.38 ± 0.82 | 91.90 ± 0.59 | 94.81 ± 0.17 |
| | Stroke | 86.51 ± 1.40 | 82.38 ± 0.82 | 93.08 ± 0.77 | 84.39 ± 1.07 | 94.81 ± 0.17 |
| | Macro-Average | 88.63 ± 0.91 | 87.73 ± 0.77 | 87.73 ± 0.77 | 88.15 ± 0.83 | 94.81 ± 0.17 |
| **DenseNet201 + ViT** | No Stroke | 90.62 ± 0.45 | 94.10 ± 1.60 | 81.90 ± 0.82 | 92.32 ± 0.93 | 94.78 ± 0.26 |
| | Stroke | 88.26 ± 2.83 | 81.90 ± 0.82 | 94.10 ± 1.60 | 84.95 ± 1.51 | 94.78 ± 0.26 |
| | Macro-Average | 89.44 ± 1.56 | 88.00 ± 1.00 | 88.00 ± 1.00 | 88.64 ± 1.22 | 94.78 ± 0.26 |
| **EfficientNet-b0 + ViT** | No Stroke | 91.21 ± 0.40 | 93.08 ± 0.00 | 83.33 ± 0.82 | 92.13 ± 0.20 | 94.44 ± 0.26 |
| | Stroke | 86.63 ± 0.11 | 83.33 ± 0.82 | 93.08 ± 0.00 | 84.95 ± 0.48 | 94.44 ± 0.26 |
| | Macro-Average | 88.92 ± 0.26 | 88.21 ± 0.41 | 88.21 ± 0.41 | 88.54 ± 0.34 | 94.44 ± 0.26 |
| **InceptionResNetV2 + ViT** | No Stroke | 90.80 ± 0.43 | 93.59 ± 0.44 | 82.38 ± 0.82 | 92.17 ± 0.44 | 95.03 ± 0.26 |
| | Stroke | 87.37 ± 0.87 | 82.38 ± 0.82 | 93.59 ± 0.44 | 84.80 ± 0.85 | 95.03 ± 0.26 |
| | Macro-Average | 89.08 ± 0.65 | 87.99 ± 0.63 | 87.99 ± 0.63 | 88.49 ± 0.64 | 95.03 ± 0.26 |

The findings in Table 7 demonstrate that in the Strategy 2 scenario, where all folds from different seeds are utilized, there is no significant difference when compared to selecting the single best model (Strategy 1), and the average 88% F1-Score band is maintained. However, in the analysis of variance, the EfficientNet-b0 architecture (± 0.34 F1-Score deviation) exhibited the highest stability among folds, proving to be one of the most resilient anomaly detection networks against hardware and data population changes (domain shift).

The variance results of the second stage of the hierarchical structure tested in the multi-center data pool are presented in Table 8.

**Table 8.** Class-based performance variance and mean values in the external test set (Ischemic / Hemorrhagic) under Strategy 2 configuration

| Fusion Architecture | Class | Precision (%) | Recall (%) | Specificity (%) | F1-Score (%) | AUC (%) |
|---|---|---|---|---|---|---|
| **ResNet50 + ViT** | Ischemic | 76.47 ± 0.99 | 58.54 ± 1.43 | 68.66 ± 0.97 | 66.31 ± 1.29 | 69.20 ± 1.42 |
| | Hemorrhagic | 48.78 ± 1.21 | 68.66 ± 0.97 | 58.54 ± 1.43 | 57.04 ± 1.16 | 69.20 ± 1.42 |
| | Macro-Average | 62.62 ± 1.10 | 63.60 ± 1.19 | 63.60 ± 1.19 | 61.68 ± 1.22 | 69.20 ± 1.42 |
| **DenseNet201 + ViT** | Ischemic | 76.70 ± 0.66 | 60.59 ± 0.44 | 67.98 ± 1.05 | 67.70 ± 0.47 | 70.62 ± 0.64 |
| | Hemorrhagic | 49.78 ± 0.57 | 67.98 ± 1.05 | 60.59 ± 0.44 | 57.48 ± 0.73 | 70.62 ± 0.64 |
| | Macro-Average | 63.24 ± 0.60 | 64.29 ± 0.65 | 64.29 ± 0.65 | 62.59 ± 0.59 | 70.62 ± 0.64 |
| **EfficientNet-b0 + ViT** | Ischemic | 76.59 ± 0.77 | 59.50 ± 3.09 | 68.37 ± 1.73 | 66.94 ± 2.10 | 68.75 ± 0.99 |
| | Hemorrhagic | 49.29 ± 1.51 | 68.37 ± 1.73 | 59.50 ± 3.09 | 57.26 ± 0.95 | 68.75 ± 0.99 |
| | Macro-Average | 62.94 ± 1.04 | 63.93 ± 1.15 | 63.93 ± 1.15 | 62.10 ± 1.43 | 68.75 ± 0.99 |
| **InceptionResNetV2 + ViT** | Ischemic | 76.06 ± 0.58 | 57.68 ± 0.97 | 68.42 ± 0.51 | 65.60 ± 0.84 | 68.29 ± 0.91 |
| | Hemorrhagic | 48.17 ± 0.74 | 68.42 ± 0.51 | 57.68 ± 0.97 | 56.54 ± 0.67 | 68.29 ± 0.91 |
| | Macro-Average | 62.11 ± 0.66 | 63.05 ± 0.72 | 63.05 ± 0.72 | 61.07 ± 0.75 | 68.29 ± 0.91 |

Strategy 2 analysis revealed a critically important fact at the clinical level. Although ResNet50 showed the highest success in Strategy 1 (single best models), when the folds were combined and the ensemble decision was made, DenseNet201 + ViT regained its lead in the external dataset with an F1-Score of 62.59% and the lowest variance of ±0.59. This proves that the high stability exhibited by the DenseNet201 architecture in internal training (Figure 5) is not coincidental; it also produces the most stable (resilient) result in multi-center tests when information is combined at the fold level. The high asymmetry between the Ischemic and Hemorrhagic classes (Hemorrhage Sensitivity ~68%, Ischemia Sensitivity ~60%) persisted in Strategy 2, confirming that clinical reality perception (over-diagnosis) is an architecture-independent dataset characteristic.

## 5. Discussion

The objective of this study was to provide a comprehensive and systematic examination of the clinical generalizability (domain shift) limitations, radiological data constraints, and architectural stability weaknesses encountered by deep learning models in the process of diagnosing acute stroke from cranial CT images. The developed computer-aided diagnosis (CAD) system makes a multifaceted contribution to the literature through its preprocessing strategy, which presents radiological Hounsfield (HU) windows to networks with asymmetric data formats; its hierarchical (cascaded) classification approach, which mimics emergency department triage; and its dynamic fusion (joint cross-attention) architecture, which combines CNN and ViT backbones in a common feature space.

One of the biggest methodological illusions in the medical AI literature is measuring the structural capabilities of models solely by metrics obtained in the test segmentation of the (homogeneous) dataset on which they were trained. A review of the existing literature reveals that the StrokeNeXt model reports a 98.8% F1 score, the DeepBrain (ResNet50) system 99.72% accuracy, and various YOLOV9-based segmentation networks report over 99% mAP success [5,11,32]. However, these extremely high success metrics are a result of the model memorizing (overfitting) scanner noise and contrast settings in the homogeneous dataset, rather than learning biological pathology. Our champion architecture, which achieved a high F1 score of 95.48% in the internal training phase of our study, regressed to the 58-62% range in independent and multicenter external cohorts (AISD, BHSD, PhysioNet), objectively demonstrating the destructive nature of this 'memorization' problem and the 'domain shift' effect. A similar weakness was detected in the NadHed.AI model, one of the rare current studies with external validation data; the purely diagnostic success based on cranial CT decreased to 88.50% in the external test set, and this bottleneck could only be overcome by integrating the clinical symptom weights of the patients (facial paralysis, aphasia) into the model, increasing it to 94.20% [3]. On the other hand, the DREAM framework demonstrated strong resilience against data scarcity by achieving 89.60% zero-shot accuracy in external data with semi-supervised learning (SSL) and dual consistency algorithms [8]. However, most of these studies have either only performed two-class (stroke present/absent) validation or have been limited to anatomical regional orientations (AGTS-Net) [19]. The Mask Anchor isolation applied in our study, which eliminated label noise in the test set, allowed us to test the direct "clinical lesion detection" ability of our models in the real world and provided insight into the adequacy of existing CAD systems on their own.

Many stroke diagnostic studies (e.g., Adablanu et al. based on Xception or Çınar et al. using EfficientNet) eliminate X-ray attenuation coefficients (HU) by directly converting radiological DICOM data to standard 8-bit PNG format [12,14]. The literature clearly shows that this data loss pushes networks into a dangerous “shortcut learning” trap. In a study conducted by Koska et al. using Otsu thresholding, it was proven that when the skull bones at +1000 HU were deleted and only the pathological search space in the 0-90 HU band was left, the accuracy of a standard MobileNetV2 network suddenly jumped from 74% to 91% [4]. Aksoy et al.'s XRAI explainability analysis on ResNet-18 also supported this situation, revealing that the models focused on skull borders and device artifacts rather than the lesion. The Radiology-Focused Asymmetric Standardization proposed in our study overcomes this shortcut learning barrier with its triple HU windowing system (Brain, Hemorrhagic, Ischemic) that eliminates impossible pixels. Moreover, unlike hybrid models in the literature that use standard data formats such as StroQ-Net or EffNet-ViT, this structure establishes an asymmetric data pipeline specific to the structural needs of architectures by feeding the data to the CNN backbone in lossless matrix (.mat) format that preserves mathematical precision, and to the ViT backbone in ".png" format for spatial analysis [9,10].

In clinical practice, detecting hyperacute ischemic strokes is quite difficult, even for the human eye, due to subtle hypodensities (similarity traps) at the gray-white matter boundary. The class-based external validation results we obtained in our second hierarchical stage (Ischemia/Hemorrhage) revealed that artificial intelligence is also deeply affected by this ambiguity. All fusion models tested exhibited relatively high precision (~74-77%) but very low sensitivity (~52-57%) in diagnosing ischemic stroke. In hemorrhagic stroke, the situation reversed, with sensitivity rising to 71% while precision dropped to 45-49%. This "diagnostic asymmetry" clearly demonstrates that the models interpret any brightness in external data (device noise, bone reflection, calcification) as hemorrhage, leading to "overdiagnosis" (false positives); however, they behave very conservatively when detecting ischemic (dark/hypodense) tissues, missing cases (false negatives). Although Adablanu et al. successfully attempted to prevent these ischemic leaks (false negatives) using Focal Loss in the literature [14], our study shows that domain shift caused by different scanner noises is a structural data problem that cannot be completely corrected with purely algorithmic manipulations (loss functions) [14].

For clinical validation of AI models, not only 'mean maximum accuracy' but also decision consistency is a vital factor. Satsangi and Narayan's Jensen-Shannon Confidence (JSC) Score analysis on Swin, CvT, and ResNet backbones demonstrated how the decision mechanisms of the models diverge and confidence bounds weaken in challenging lesions [21]. In this context, the multi-seed and layered external testing strategies (Strategies 1 and 2) that we designed revealed a very critical architectural behavior. In Strategy 1, where only individual champions were evaluated, ResNet50 + ViT, which extracts deeper hierarchical features, appeared as the most resilient network to external datasets (domain shift), while in Strategy 2, where all k-layered models within the seed made a joint decision (soft voting), the DenseNet201 + ViT architecture ranked first with a 62.59% F1 score and the lowest volatility (variance) of ±0.59. This result proves that densely connected feature maps in the DenseNet structure are more resistant to over-learning and that the model does not lose its structural stability in its decisions (high structural resilience) when faced with heterogeneous data obtained from different hardware (devices).

### 5.1 Limitations

Although this study provides transparent metrics to the literature in external validation and architectural variance testing, it has certain limitations. First, the models analyze 2D axial sections, ignoring the 3D volumetric integrity (Z-axis spatial relationships) of the cranial CT structure. Second, as highlighted in the NadHed.AI study, in clinical practice, radiological images are examined in conjunction with patient history (symptoms, age, time of presentation) [3]. A notable limitation of this study is the absence of multimodal patient data within the utilized datasets. Our future goal is to integrate the multi-channel fusion architecture we designed with structural patient data (EHR) by directly converting it to a multimodal structure, as well as to optimize (Bayesian approaches) different CNN models using feature extractors to better fine-tune them to the problem under study.

## 6. Conclusion

In this study, a hierarchical computer-aided diagnosis (CAD) system was developed that can directly mimic clinical workflows in acute stroke management, supported by asymmetric data standardization and a joint cross-attention architecture. The developed Radiology-Focused Multichannel Standardization pipeline successfully prevented the shortcut learning problem frequently encountered in the literature by isolating ischemic, hemorrhagic, and normal tissue windows with mathematical limits. However, the greatest contribution of our study to the literature is that it goes beyond the high laboratory success reported with a single dataset and reveals the "domain shift" vulnerabilities (true generalizability limits) of the models in the external world. Multicenter tests (AISD, BHSD, PhysioNet) have shown that CAD systems experience diagnostic asymmetry in the face of artifacts from different tomography scanners. It has been shown to operate with very low sensitivity (57%) while producing high precision (77%) in ischemic cases, and is prone to overdiagnosis in suspected bleeding. Within these challenging clinical boundaries, the DenseNet201+ViT hybrid architecture, which showed the lowest deviation (±0.59) and highest stability (62.59% F1-Score) in Strategy-2 K-Fold Ensemble analysis of variance, emerged as the most reliable backbone for clinical integration across different hardware and patient populations. Future studies should focus on models that can be better fine-tuned with three-dimensional volumetric feature extraction and direct integration of patient clinical symptom weights into the model (multimodality) to break the diagnostic asymmetry of this hierarchical architecture.

## Declarations

**Supplementary Materials:** Not applicable.

**Author contributions:** T.U. conceived the idea; T.U. curated the data; T.U. analyzed the data; and T.U. led the writing; T.U. revised and edited the manuscript.

**Funding:** This research received no external funding.

**Institutional Review Board Statement:** Not applicable.

**Informed Consent Statement:** Not applicable.

**Data Availability Statement:** Dataset available in a publicly accessible repository. The original dataset used in the study are openly available in Republic of Türkiye Ministry of Health, Open Data Portal at https://acikveri.saglik.gov.tr/Home/DataSetDetail/1. Research data available on request due to restrictions. All research data has been uploaded to Zenodo (DOI:

10.5281/zenodo.22132268). The data presented in this study are available on request from the corresponding author due to ongoing research studies.

**Acknowledgments:** Not applicable.

**Conflicts of Interest:** The authors declare no conflicts of interest.

**Abbreviations**

AGTS-Net: Anatomically Guided Two-Stage Network.

AISD: Acute Ischemic Stroke Dataset.

AUC: Area Under the Curve.

BHSD: Brain Hemorrhage Segmentation Dataset.

CAD: Computer-aided diagnostic.

CNN: Convolutional Neural Network.

CT: Computed tomography / Cranial Computed Tomography.

EHR: Structural patient data.

FN: False Negative.

FP: False Positive.

HU: Hounsfield Unit.

NCCT: Non-contrast CT.

SSL: Semi-supervised learning.

TN: True Negative.

TP: True Positive.

tPA: Tissue plasminogen activator.

ViT: Vision Transformer.